\makeatletter\def\graphicscache@inhibit{true}\makeatother

\documentclass[hidelinks, letterpaper, 10 pt, conference]{ieeeconf}  %

\IEEEoverridecommandlockouts                              %

\usepackage{graphicx}
\usepackage{color}
\usepackage{comment}
\usepackage{subfigure}
\usepackage{amssymb}
\usepackage[export]{adjustbox}
\usepackage{xspace}
\def\ie{i.e.\xspace}
\def\eg{e.g.\xspace}
\def\etal{et~al.\xspace}
\usepackage[normalem]{ulem}
\usepackage{graphicscache}

\usepackage{tikz}
\usetikzlibrary{arrows}
\usetikzlibrary{arrows.meta}
\usetikzlibrary{positioning,calc}
\usetikzlibrary{decorations.pathreplacing}
\usetikzlibrary{decorations.markings}
\usetikzlibrary{fit}
\usetikzlibrary{shapes.callouts}
\usetikzlibrary{shapes.geometric}
\usetikzlibrary{matrix}
\usepackage{pgfplots}
\usepgfplotslibrary{statistics}
\pgfplotsset{compat=newest}

\usepackage{url}

\title{\LARGE \bf
A VR System for Immersive Teleoperation and Live Exploration\\with a Mobile Robot}

\author{Patrick Stotko$^{1}$, Stefan Krumpen$^{1}$, Max Schwarz$^{2}$, Christian Lenz$^{2}$,\\Sven Behnke$^{2}$, Reinhard Klein$^{1}$, and Michael Weinmann$^{1}$%
\thanks{$^{1}$P. Stotko, S. Krumpen, R. Klein, and M. Weinmann are with the Institute of Computer Science II -- Computer Graphics,
        University of Bonn, Germany
        {\tt\small \{stotko,krumpen,rk,mw\}@cs.uni-bonn.de}}%
\thanks{$^{2}$M. Schwarz, C. Lenz and S. Behnke are with the Institute of Computer Science VI -- Autonomous Intelligent Systems,
        University of Bonn, Germany
        {\tt\small \{schwarz,lenz\}@ais.uni-bonn.de, behnke@cs.uni-bonn.de}}%
\thanks{This work was supported by the DFG projects KL 1142/11-1 and BE 2556/16-1 (DFG Research Unit FOR 2535 Anticipating Human Behavior) as well as KL 1142/9-2 and BE 2556/7-2 (DFG Research Unit FOR 1505 Mapping on Demand).}%
}

\begin{document}

\maketitle
\thispagestyle{empty}
\pagestyle{empty}

\begin{abstract}
\begin{tikzpicture}[remember picture,overlay]%
\node[anchor=south,align=center,font=\mdseries,yshift=0.3cm, xshift=-0.3cm] at (current page.south) {%
\begin{minipage}[c]{\textwidth}\copyright \ 2019 IEEE. Personal use of this material is permitted. Permission from IEEE must be obtained for all other uses, in any current or future media, including reprinting/republishing this material for advertising or promotional purposes,creating new collective works, for resale or redistribution to servers or lists, or reuse of any copyrighted component of this work in other works. The final version of record is available at \url{http://dx.doi.org/10.1109/IROS40897.2019.8968598}.\end{minipage}%
};%
\end{tikzpicture}%
Applications like disaster management and industrial inspection often require experts to enter contaminated places.
To circumvent the need for physical presence, it is desirable to generate a fully immersive individual live teleoperation experience.
However, standard video-based approaches suffer from a limited degree of immersion and situation awareness due to the restriction to the camera view, which impacts the navigation.
In this paper, we present a novel VR-based practical system for immersive robot teleoperation and scene exploration.
While being operated through the scene, a robot captures RGB-D data that is streamed to a SLAM-based live multi-client telepresence system.
Here, a global 3D model of the already captured scene parts is reconstructed and streamed to the individual remote user clients where the rendering for \eg head-mounted display devices (HMDs) is performed.
We introduce a novel lightweight robot client component which transmits robot-specific data and enables a quick integration into existing robotic systems.
This way, in contrast to first-person exploration systems, the operators can explore and navigate in the remote site completely independent of the current position and view of the capturing robot, complementing traditional input devices for teleoperation.
We provide a proof-of-concept implementation and demonstrate the capabilities as well as the performance of our system regarding interactive object measurements and bandwidth-efficient data streaming and visualization.
Furthermore, we show its benefits over purely video-based teleoperation in a user study revealing a higher degree of situation awareness and a more precise navigation in challenging environments.

\end{abstract}

\section{Introduction}

\begin{figure}[t]
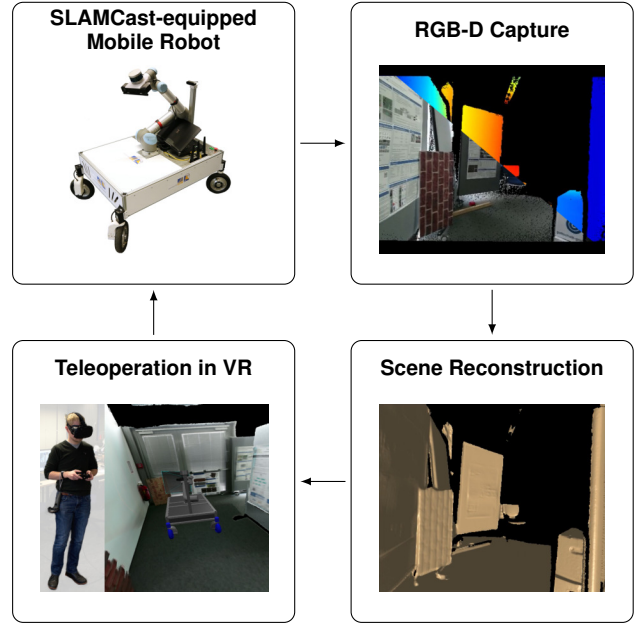

    \centering
﻿\begin{tikzpicture}[
	scale=0.75,
 	font=\sffamily\footnotesize,
    every node/.append style={text depth=.2ex},
	bigbox/.style={rectangle,rounded corners,draw=black,align=center},
	smallbox/.style={rectangle,rounded corners,draw=black,align=center, font=\sffamily\scriptsize},
	label/.style={font=\sffamily\footnotesize,align=center},
]

\draw[rounded corners] (0,0) rectangle ++(5,5);
\node[label] at (2.5,4.5) {\textbf{Teleoperation in VR}};

\draw[rounded corners] (6,0) rectangle ++(5,5);
\node[label] at (8.5,4.5) {\textbf{Scene Reconstruction}};

\draw[rounded corners] (6,6) rectangle ++(5,5);
\node[label] at (8.5,10.5) {\textbf{RGB-D Capture}};

\draw[rounded corners] (0,6) rectangle ++(5,5);
\node[label] at (2.5,10.5) {\textbf{SLAMCast-equipped}\\\textbf{Mobile Robot}};

\node[anchor=south west,inner sep=0,outer sep=0] at (1,6.5) {\includegraphics[height=2.5cm]{figures/mario.png}};
\node[anchor=south west,inner sep=0,outer sep=0] at (6.5,0.5) {\includegraphics[width=3cm]{figures/reconstruction_teaser.png}};
\node[anchor=south west,inner sep=0,outer sep=0] at (0.5,0.5) {\includegraphics[width=3.03cm]{figures/operator_teaser.png}};    %
\node[anchor=south west,inner sep=0,outer sep=0] at (6.5,6.5) {\includegraphics[width=3cm]{figures/raw_data_teaser.png}};

\draw[-latex] (5.1,8.5)- ++(0.8,0) ;
\draw[-latex] (8.5,5.9)- ++(0,-0.8) ;
\draw[-latex] (5.9,2.5)- ++(-0.8,0) ;
\draw[-latex] (2.5,5.1)- ++(0,0.8) ;

\end{tikzpicture}
     \caption{High-level overview of our novel immersive robot teleoperation and scene exploration system where an operator controls a robot using a live captured and reconstructed 3D model of the environment.}
    \label{fig:teaser}
\end{figure}

Due to the significant progress in VR displays in recent years, the immersive exploration of scenes based on virtual reality systems has gained a lot of attention with diverse applications in entertainment, teleconferencing~\cite{Orts-Escolano:2016}, remote collaboration~\cite{stotko2019slamcast}, medical rehabilitation and education.
The quality of immersive experience of places, while being physically located in another environment, opens new opportunities for robotic teleoperation scenarios.
Here, the major challenges include aspects such as resolution and frame rates of the involved display devices or the presentation and consistency of the respective data that increase the awareness of being immersed into the respective scene~\cite{Fontaine:1992,Held:1992,Witmer:1998}.
Another key challenge is the preservation of a high degree of situation awareness regarding the teleoperated robot's pose within its physical environment to allow precise navigation.

Purely video-based robot teleoperation and scene exploration is rather limited in the sense that the view is directly coupled to the area observed by the camera.
This affects/impacts both the degree of immersion and the degree of situation awareness as remotely maneuvering a robot without having a complete overview regarding its current local environment is challenging, especially in case of narrow doors or corridors.
Furthermore, remembering the locations of relevant scene entities is also complicated for video-only teleoperation which impacts independent visual navigation to already previously observed scene parts outside the current camera view.
In contrast, transmitting the scene in terms of a reconstructed 3D model and immersing the teleoperator into this virtual scene is a promising approach to overcome these problems.
Highly efficient real-time 3D reconstruction and real-time data transmission recently have been proven to be the key drivers to high-quality tele-conferencing within room-scale environments~\cite{Orts-Escolano:2016} or for immersive telepresence based remote collaboration tasks in large-scale environments~\cite{stotko2019slamcast}.
The benefit regarding situation awareness can still be preserved in case of network interruptions as the remote user remains immersed into the so far reconstructed scene and, after re-connection, newly arriving data can directly be integrated into the already existing scene model.
However, a manual capturing process as used by Stotko \etal~\cite{stotko2019slamcast} is not possible within contaminated places.
To the best of our knowledge, these kind of systems have not been adapted to the constraints of robot teleoperation -- in our opinion, because the quality and scalability of 3D reconstruction methods has been too low until recently.

In this paper, we tackle the aforementioned challenges based on a novel system for immersive robot teleoperation and scene exploration within live-captured environments for remote users based on virtual reality and real-time 3D scene capture (see Fig.~\ref{fig:teaser}).
This creation of an immersive teleoperation experience implies that the aforementioned conditions are met under strong time constraints to allow an immersive live teleoperation of the robot within the considered scenes and, hence, relies on on-the-fly scene reconstruction, immediate data transmission and visualization of the models to remote-connected users.
For this purpose, our system involves a robot which is teleoperated through a respective scenario while capturing RGB-D data.
To provide an as-complete-as-possible scene reconstruction for the teleoperation, the involved RGB-D camera can be moved via a manipulator, if existing on the robot.
The captured data is sent to a reconstruction client component, that performs real-time dense volumetric Simultaneous Localization And Mapping (SLAM) based on voxel block hashing, and the current 3D model is managed on the server based on an efficient hash map data structure.
Finally, the current model is streamed to the remote exploration client based on a low-bandwidth representation.
Our approach allows a re-thinking regarding current exploration scenarios as encountered in \eg disaster management, so that, on the long term, humans do not have to be exposed to \eg contaminated environments but still can interact with the environment.
It is furthermore desirable to add the functionality offered by the proposed framework to existing robotic systems.
Therefore, we impose no requirements on the robotic platform:
The robot-side system, consisting of an RGB-D camera and a notebook, is entirely self-contained.
Optional interfaces allow tighter integration with the robot.
Besides an evaluation of the performance of our system in terms of bandwidth requirements, visual quality and overall lag, we additionally provide the results of a psychophysical study that indicates the benefit of immersive VR based teleoperation in comparison to purely video-based teleoperation.
Finally, we also show several example applications by demonstrating how the remote users can interact with both the robot and the scene.

In summary, the main contributions of this work are:
\begin{itemize}
    \item The development of a novel system for immersive robot teleoperation and scene exploration within live-captured environments for remote users based on virtual reality and fast 3D scene capture -- as needed \eg for the inspection of contaminated scenes that cannot directly be accessed by humans,
    \item the implementation of the aforementioned system in terms of hardware and software,
    \item the evaluation of the benefits offered by this kind of immersive VR-based robot teleoperation over purely video-based teleoperation in the scope of a respective psychophysical study, and
    \item the evaluation of the system within proof-of-concept experiments regarding the robotic application of remote live site exploration.
\end{itemize}

\section{Related Work}

In this section, we review the progress made in telepresence systems with a particular focus on their application for teleoperation and remote collaboration involving robots.

\paragraph*{Telepresence Systems}

The key to success for the generation of an immersive and interactive telepresence experience is the real-time 3D reconstruction of the scene of interest.
In particular due to the high computational burden and the huge memory requirements required to process and store large scenes, seminal work on multi-camera telepresence systems~\cite{Fuchs:1994,Kanade:1997,Mulligan:2000,Towles:2002,Tanikawa:2005,Kurillo:2008} with less powerful hardware available at that time faced limitations regarding the capability to capture high-quality 3D models in real-time and to immediately transmit them to remote users.
More recently, the emerging progress towards affordable commodity depth sensors including \eg the Microsoft Kinect has successfully been exploited for the development of 3D reconstruction approaches working at room scale~\cite{Maimone:2012,Maimone:2012b,Molyneaux:2012,Jones:2014}.
Yet the step towards high-quality reconstructions remained highly challenging due to the high sensor noise as well as temporal inconsistency in the reconstructed data.

Recently, a huge step towards an immersive teleconferencing experience has been achieved with the development of the Holoportation system~\cite{Orts-Escolano:2016}.
This system has been implemented based on the Fusion4D framework~\cite{Dou:2016} that allows an accurate 3D reconstruction at real-time rates, as well real-time data transmission and the coupling to AR/VR technology.
However, real-time performance is achieved based on massive hardware requirements involving several high-end GPUs running on multiple desktop computers and most of the hardware components have to be installed at the local user's side.
Furthermore, only an area of limited size that is surrounded by the involved static cameras can be captured which allows the application of this framework for teleconferencing but prevents it from being used for interactive remote exploration of larger live-captured scenes.

Towards the goal of exploring larger environments as related to the exploration of contaminated scenes envisioned in this work, Mossel and Kr{\"o}ter~\cite{mossel} presented a system that allows interactive VR-based exploration of the captured scene by a single exploration client.
Their system benefits from the real-time reconstruction based on current voxel block hashing techniques~\cite{infinitam}, however, it only allows scene exploration by one single exploration client, and, yet, the bandwidth requirements of this approach have been reported to be up to 175\,MBit/s.
Furthermore, the system relies on the direct transmission of the captured data to the rendering client, which is not designed to handle network interruptions that force the exploration client to reconnect to the reconstruction client and, consequently, scene parts that have been reconstructed during network outage will be lost.

The recent approach by Stotko \etal~\cite{stotko2019slamcast} overcomes these problems and allows the on-the-fly scene inspection and interaction by an arbitrary number of exploration clients, and, hence, represents a practical framework for interactive collaboration purposes.
Most notably, the system is based on a novel compact Marching Cubes (MC) based voxel block representation maintained on a server.
Efficient streaming at low-bandwidth requirements is achieved by transmitting MC indices and reconstructing and storing the models explored by individual exploration clients directly on their hardware.
This makes the approach both scalable to many-client-exploration and robust to network interruptions as the consistent model is generated on the server and the updates are streamed once the connection is re-established.

\paragraph*{Robot-based Remote Telepresence}

The benefits of an immersive telepresence experience have also been investigated in robotic applications.
Communication via telepresence robots (\eg\cite{ Kristoffersson:2013,Rae:2014,Yang:2017}) is typically achieved based on a video/audio communication unit on the robot.
More closely related to our approach are the developments regarding teleoperation in the context of exploring scenes.
Here, remote users usually observe the video stream acquired by the cameras of the involved exploration robots to perform \eg the navigation of the robot though a scene as well as the inspection of certain objects or areas.
The visualization can be performed based on projecting live imagery onto large screens~\cite{WETTERGREEN1999}, walls~\cite{Roberts:2015}, monitors~\cite{Podnar:2006, Rekleitis:2010,Macharet:2012,KaptelininBDW17} or based on head-mounted display (HMD) devices~\cite{Hine:1994,Elliott:2012,Martinez-Hernandez:2015, Martinez-Hernandez:2017,Peppoloni:2015,Kurup:2016,Lipton:2018}.
Some of this work~\cite{Martinez-Hernandez:2015,Peppoloni:2015,Kurup:2016, Lipton:2018} additionally coupled the interactions recorded by the HMD device to perform a VR-based teleoperation.
However, the dependency on the current view of the used cameras does not allow an independent exploration of the scene required \eg when remote users with different expertise have to focus on their individual tasks.
Most closely related to our work is the approach of Bruder \etal~\cite{Bruder:2014}, where a point cloud based 3D model of the environment is captured by a mobile robot and displayed by a VR-HMD.
As discussed by the authors, the sparsity of the point cloud leads to the impression that objects or walls only appear solid when being observed from a sufficient distance and dissolve when being approached.
This distance, in turn, also depends on the density of the point cloud.
Furthermore, common operations including selection, manipulation, or deformation have to be adapted as ray-based approaches cannot be applied.
Our approach overcomes these problems by capturing a surface-based 3D mesh model that can be immersively explored via live-telepresence based on HMDs.

\paragraph*{Robot Platform}

In Schwarz \etal~\cite{Schwarz:2018}, the rescue robot Momaro is described, which is equipped with interfaces for immersive teleoperation using an HMD device and 6D trackers.
The immersive display greatly benefited the operators by increasing situational awareness.
However, visualization was limited to registered 3D point clouds, which carry no color information.
As a result, additional 2D camera images were displayed to the operator to visualize texture.
Momaro served as a precursor to the Centauro robot~\cite{Klamt:2018}, which extends the Momaro system in several directions, including immersive display of RGB-D data.
However, the system is currently limited to displaying live data without aggregation.

\section{Overview}
\label{sec:overview}

\begin{figure}
    \centering
﻿\begin{tikzpicture}[
 	font=\sffamily\footnotesize,
    every node/.append style={text depth=.2ex},
	bigbox/.style={rectangle,rounded corners,draw=black,align=center},
	smallbox/.style={rectangle,rounded corners,draw=black,align=center, font=\sffamily\scriptsize},
   robot_color/.style={fill=yellow!40},
   slam_color/.style={fill=green!40}
]

\coordinate (robo) at (0,0);
\draw[rounded corners] (robo) rectangle ++(4,6);

\draw[rounded corners, fill = yellow!20, dotted] ($(robo) + (1.75, 0.25)$) rectangle ++(2.1,2.5);
\draw[rounded corners, fill=green!20, dotted] ($(robo) + (1.75, 3)$) rectangle ++(2.1,2.5);
\node[align=center,font=\sffamily\scriptsize] at ($(robo) + (2.8,2.5)$) {Robot Computer};
\node[align=center,font=\sffamily\scriptsize] at ($(robo) + (2.8,5.2)$) {Notebook};
\node at ($(robo) + (2.0,5.75)$) {\textbf{Robot}};
\node[smallbox, slam_color] at ($(robo) + (0.75,4.5)$){RGB-D\\Sensor};
\node[smallbox, robot_color] at ($(robo) + (0.75,0.75)$){Robot\\Hardware};

\node[smallbox, robot_color] at ($(robo) + (2.8,0.75)$){Robot\\Control};
\node[smallbox,slam_color] at ($(robo) + (2.8,1.750)$){Robot\\Client};
\node[smallbox, slam_color] at ($(robo) + (2.8,3.5)$){SLAMCast\\Server};
\node[smallbox, slam_color] at ($(robo) + (2.8,4.5)$){Reconstruction\\Client};

\coordinate (teleop) at (6,0);
\draw[rounded corners] (teleop) rectangle ++(2.4,6);
\node at ($(teleop) + (1.2,5.75)$) [align=center]{\textbf{Operator}};
\node[smallbox, slam_color] at ($(teleop) + (1.2,3.5)$) {Exploration\\Client};
\node[smallbox, slam_color] at ($(teleop) + (1.2,4.5)$) {Head-mounted\\Display};

\node[smallbox, robot_color] at ($(teleop) + (1.2,0.75)$) {Teleoperation\\Interface};

\draw[-latex] ($(robo) +(1.3,4.5)$) - ++(0.5,0);
\draw[-latex] ($(robo) +(2.2,0.75)$) - ++(-0.75,0);
\draw[-latex] ($(robo) +(2.8,4.1)$) - ++(0,-0.2);
\draw[-latex] ($(teleop) +(1.2,3.9)$) - ++(0,0.2);
\draw[-latex] ($(robo) +(3.6,3.5)$) - ++(2.75,0);
\draw[-latex] ($(robo) +(6.2,0.75)$) - ++(-2.75,0);
\draw[-latex] ($(robo) +(2.2,1.75)$) - |++(-0.75,0) -| ++ (0,1.75) - ++(0.55,0);

\draw[draw=black, dashed](5,0.5)-- ++ (0,5);
\node at (5,5.75) {\textbf{WiFi}};

\end{tikzpicture}

     \caption{Implementation of our immersive teleoperation system. The system allows the operator to immerse into the reconstructed scene to gain a third-person overview, while teleoperating the robot using existing teleoperation devices (\eg a gamepad). Components in green are part of the SLAMCast framework; yellow boxes correspond to existing parts of the robotic system.}
    \label{fig:workflow}
\end{figure}
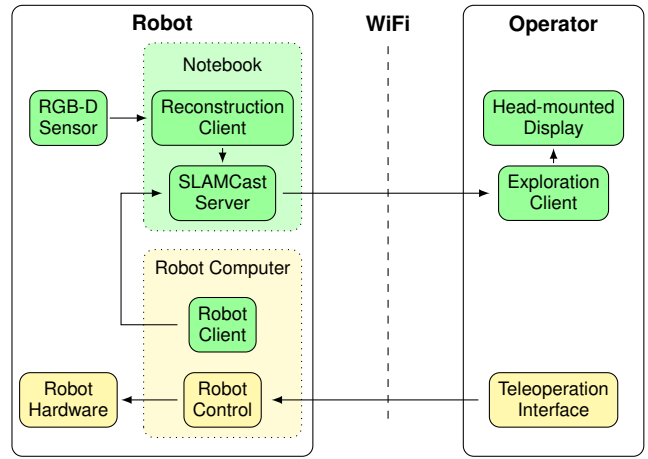

The main goal of this work is the design and implementation of a practical system for immersive robot teleoperation and scene exploration within live-captured environments for remote users based on virtual reality and real-time 3D scene capture (see Fig.~\ref{fig:workflow}).
For this purpose, our proposed system involves (1) a robotic platform moving through the scene and performing scene capture, (2) an optional robot client that provides information about the current robot posture, (3) a reconstruction client that takes the captured data and computes a 3D model of the already observed scene parts, (4) a server that maintains the model and controls the streaming to the individual exploration clients, and (5) the connected exploration clients that perform the rendering \eg on HMDs and can be used for teleoperation.
By design, our system offers the benefits of allowing a large number of exploration clients, where, in addition to the teleoperator maneuvering the robot, several remote users may independently inspect the reconstructed scene and communicate with each other, \eg for disaster management purposes.

In the following, we provide more details regarding the implementation of the involved components.

\section{Robot-based Scene Scanning}
\label{sec:robot}

\begin{figure}[t]
    \centering
﻿\begin{tikzpicture}[
	scale=0.75,
 	font=\sffamily\footnotesize,
    every node/.append style={text depth=.2ex},
	bigbox/.style={rectangle,rounded corners,draw=black,align=center},
	smallbox/.style={rectangle,rounded corners,draw=black,align=center, font=\sffamily\scriptsize},
]

\node[anchor=south west,inner sep=0,outer sep=0] at (0,0) {\includegraphics[width=4.5cm]{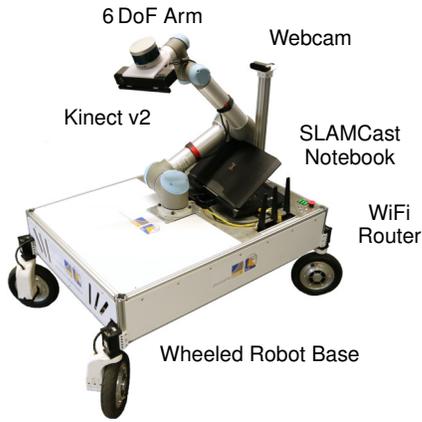}};

\node at (2.6,7.2) {6\,DoF Arm};
\node at (1.8,5.4) {Kinect v2};
\node at (4.5,1.2) {Wheeled Robot Base};
\node at (5.4,6.8) {Webcam};
\node[align = center] at (6.1,4.9) {SLAMCast\\Notebook};
\node[align = center] at (6.8,3.5) {WiFi\\Router};

\end{tikzpicture}
     \caption{The Mario robot is the exemplary target platform of our work. It has been equipped with an additional Kinect v2 RGB-D sensor and a notebook for processing and streaming of the reconstructed scene.}
    \label{fig:mario}
\end{figure}

Mobile scene scanning was performed using the ground robot Mario (see Fig.~\ref{fig:mario}), a robot with steerable wheels capable of omnidirectional locomotion.
Mario won the Mohammed bin Zayed International Robotics Challenge 2017 (MBZIRC)\footnote{\url{http://www.mbzirc.com}} both in the UGV task and the Grand Challenge.
For details on Mario, we refer to the work of Schwarz \etal\cite{Schwarz2018}.
Important for this work, Mario offers a large footprint, which yields high stability and few high-frequency movements of the camera.
On the other hand, Mario can be difficult to maneuver in tight spaces, since it is designed for high-speed outdoor usage.
Mario can be operated remotely using a WiFi link based on various sensors on the robot.

The key features of the robotic capturing system are:

\paragraph*{Driving Unit}

Based on the assumption of mostly flat terrain, we used a four-wheel-based robot system to allow a stable operation.
In particular, we use an omnidirectional base due to its benefits regarding the precise positioning of the robot and the avoidance of complicated maneuvering for small adjustments as required in our envisioned contaminated site exploration scenario.
Driven by the requirements of MBZIRC, the direct-drive brushless DC hub motors inside each steerable wheel allow reaching velocities of up to 4\,m/s.
In the indoor exploration scenario considered here, we limit the velocity to 0.15\,m/s.

\paragraph*{Robot Arm}

Mario is equipped with a Universal Robots UR5, an off-the-shelf 6 DoF arm which offers more than sufficient working range to pan and tilt the endeffector-mounted camera sensor in order to increase the captured scene area.
During scene exploration, the camera is automatically moved along Z-shaped trajectories to increase the field of view and thus the completeness of the captured model.

\paragraph*{RGB-D Sensor}

We extended the arm with the Microsoft Kinect v2, an off-the-shelf RGB-D sensor.
This camera provides RGB-D data with a resolution of 512$\times$424 pixels at 30\,Hz.
Note that RGB-D sensors in smartphones like the ASUS Zenfone AR sensor could also be used.
Although these have a lower resolution and frame rate, they still allow for a sufficient reconstruction as shown by Stotko \etal~\cite{stotko2019slamcast}.

\paragraph*{Electrical System}

To meet the high voltage requirements imposed by the brushless wheel motors, the robot is powered by an eight-cell LiPo battery with 16\,Ah and 29.6\,V nominal voltage which allows operation times of up to 1\,h depending on the task intensity.
The UR5 arm is also run directly from the battery.

\paragraph*{Data Transmission}

The system is equipped with a Netgear Nighthawk AC1900 router that allows remotely monitoring the system as well as transmission of the scene data to clients.
Additionally, the robot is equipped with a Velodyne VLP-16 3D LiDAR as well as a wide-angle Logitech webcam (that can be used for teleoperation).
To keep requirements minimal, we did not integrate the LiDAR into our system, although this is a possible extension point.
During the experiments, the robot is teleoperated through an existing wireless gamepad interface, which controls the omnidirectional velocity (2D translation and rotation around the vertical axis).
We do not impose any requirements on the teleoperation method besides that it is compatible with third-person control, \ie that it is usable while standing next to the robot (in reality or in VR).

\section{Live Teleoperation and Exploration System}
\label{sec:telepresence}

The aforementioned robotic capturing system is used in combination with an efficient teleoperation system consisting of the following components:

\subsection{Reconstruction Client}

RGB-D data captured by the robot are transmitted to the reconstruction client component, where a dense virtual 3D model is reconstructed in real-time using volumetric fusion into a sparse set of spatially-hashed voxel blocks based on implicit truncated signed distance fields (TSDFs)~\cite{niessner,infinitam}.
Fully reconstructed voxel blocks, \ie blocks that fall outside the current camera frustum, are queued for transmission to the central server component.
Furthermore, the set of actively reconstructed visible voxel blocks is also added to the set of to-be-streamed blocks when the robot stops moving as well as at the end of the session~\cite{stotko2019slamcast}.
Subsets of these blocks are then progressively fetched, compressed using lossless real-time compression~\cite{Collet:2017:zstd}, and streamed to the server.
In addition, the reconstruction client transmits the current estimated camera pose to the server which is broadcasted to the exploration clients and used for the visualization of the camera's view frustum and the robot within the scene.

\subsection{Robot Client}

We introduce a novel component in the SLAMCast framework that allows the efficient and modular extension to a robot-based live telepresence and teleoperation system.
This component is required if the camera is actuated on the robot -- in this case, the pose of the robot components cannot be computed from the camera pose alone.
The robot client solves this problem by providing the SLAMCast system with the poses of all robot links (in our exemplary case with Mario the posture of the 6\,DoF arm as well as the wheel orientations).
This information is transmitted to the SLAMCast server and then broadcasted to the exploration clients.
In combination with the estimated camera pose, this enables an immersive visualization of the robot within the scene.
Note that the interface to the robotic system could be extended by streaming additional sensor data (\eg LiDAR data) to the server.
However, this work focuses on a minimally-invasive solution for immersive teleoperation and such extensions are thus out of scope.

\subsection{Server}

The server component manages the global model as well as the stream states of each connected exploration client, \ie the set of updated voxel blocks that need to be streamed to the individual client.
For efficient streaming to the clients, the received TSDF voxel blocks are converted to the bandwidth-efficient MC voxel block representation~\cite{stotko2019slamcast} and then added to the stream sets of each connected exploration client.
Here, we used a simplified version of the Marching Cubes (MC) technique~\cite{Lorensen:1987:MCH} where the weights have been discarded.
In case a client re-connects to the server, the complete list of voxel blocks is added to its stream set in case the previously streamed parts are lost caused by \eg accidentally closing the client by the user.

\subsection{Exploration Client}

At the remote expert's site, the exploration client requests updated scene parts either based on its current viewing pose, \ie the parts that the user is currently exploring and interested in, in the order of the reconstruction, which resembles the movement of the robot, or in an arbitrary order which can be used to prefetch the remaining parts of the model outside the current view.
Once the requested compressed MC voxel data arrived, they are uncompressed and passed to a reconstruction thread which generates a triangle mesh using Marching Cubes~\cite{Lorensen:1987:MCH} as well as three additional levels of detail for efficient rendering.
Furthermore, a virtual model of the robot is visualized within the scene using the estimated camera pose as well as the poses of the robot components.
Since the estimated robot position might be affected by jittering effects due to imperfect camera poses, we apply a temporal low-pass filter on the robot's base pose.
This ensures a smooth and immersive teleoperation experience.

In addition, our system can handle changes in the scene over time as \eg occurring when doors have been opened or objects/obstacles have been removed.
This is achieved by a reset function with which the exploration client may request scene updates for selected regions.
In this case, the already reconstructed parts of the 3D model of the scene that are currently visible are deleted and the respective list of blocks is propagated to the server and exploration clients.

\section{Experimental Results}

\begin{figure}
    \centering
    \includegraphics[width=\linewidth]{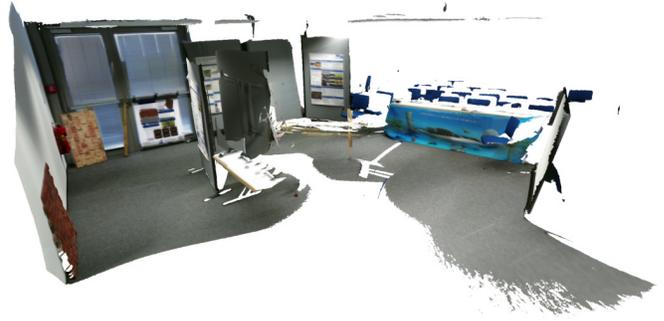}
    \caption{Reconstructed 3D model of the teleoperation scene.}
    \label{fig:teleop_reconstruction}
\end{figure}

\begin{figure}
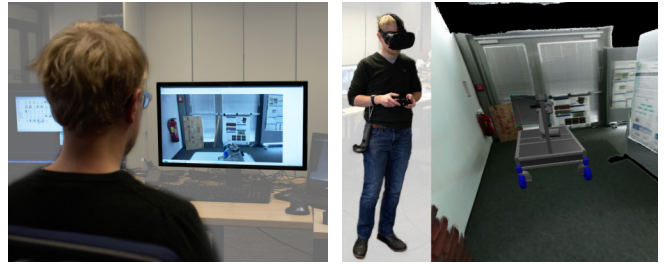

    \centering
    \includegraphics[height=.4\linewidth,clip,trim=0 0 130 0]{figures/teleop_cam.jpg}
    \hfill
    \includegraphics[height=.4\linewidth]{figures/operator_teaser.png}
    \caption{Teleoperation experiment. Left: Baseline experiment with wide-angle camera feed. Right: Teleoperation using the proposed VR system.}
    \label{fig:teleop_experiment}
\end{figure}

\begin{figure*}
    \centering
    \adjustbox{valign=T}{%
﻿\begin{tikzpicture}[font = \footnotesize, every mark/.append style={mark size=0.5pt}]
\begin{axis}[
    name=plot,
    boxplot/draw direction=x,
    width=0.6\textwidth,
    height=6.5cm,
    boxplot={
        draw position={1.5 - 0.325/2 + 1.0*floor((\plotnumofactualtype + 0.001)/2) + 0.325*mod((\plotnumofactualtype + 0.001),2)},
        box extend=0.275,
        average=auto,
        every average/.style={/tikz/mark=x, mark size=1.5, mark options=black},
        every box/.style={draw, line width=0.5pt, fill=.!40!white},
        every median/.style={line width=2.0pt},
        every whisker/.style={dashed},
    },
    ymin=1,
    ymax=13,
    y dir=reverse,
    ytick={1,2,...,13},
    y tick label as interval,
    yticklabels={
        Maintaining situation awareness,
        Resolution,
        Adequacy for teleoperation,
        Ease of controlling view,
        Robot control latency,
        Overall ease of use,
        Avoiding obstacles,
        Assessing terrain for navigability,
        Moving robot in desired position,
        Maneuvering around corners,
        Localization in the scene,
        Moving quickly
    },
    y tick label style={
        align=center
    },
    xmin=0.75,
    xmax=7.25,
    xtick={1, 2 ,..., 7},
    xticklabels = {1, 2, ..., 7},
    cycle list={{orange!50!red,green!50!black}},
    y dir=reverse,
    legend image code/.code={
        \draw [#1, fill=.!40!white] (0cm,-1.5pt) rectangle (0.3cm,1.5pt);
    },
    legend style={
        anchor=north west,
        at={($(0.0,1.0)+(0.2cm,-0.1cm)$)},
    },
    legend cell align={left},
]

\addplot
table[row sep=\\,y index=0] {
data\\
4\\3\\3\\3\\7\\6\\2\\7\\4\\7\\3\\4\\7\\4\\3\\6\\3\\5\\6\\5\\
};
\addplot
table[row sep=\\,y index=0] {
data\\
5\\6\\7\\6\\4\\6\\7\\7\\5\\7\\6\\5\\6\\4\\6\\6\\6\\4\\6\\5\\
};

\addplot
table[row sep=\\,y index=0] {
data\\
5\\5\\6\\5\\6\\3\\7\\7\\5\\5\\7\\5\\5\\4\\3\\5\\6\\6\\6\\5\\
};
\addplot
table[row sep=\\,y index=0] {
data\\
4\\3\\5\\6\\7\\4\\6\\6\\5\\5\\7\\5\\6\\4\\5\\5\\4\\4\\5\\6\\
};

\addplot
table[row sep=\\,y index=0] {
data\\
4\\4\\4\\3\\\\5\\3\\5\\5\\6\\6\\4\\6\\3\\2\\5\\4\\6\\6\\5\\
};
\addplot
table[row sep=\\,y index=0] {
data\\
5\\5\\6\\6\\6\\5\\7\\6\\5\\3\\6\\5\\7\\3\\6\\7\\5\\5\\6\\6\\
};

\addplot
table[row sep=\\,y index=0] {
data\\
4\\3\\3\\4\\4\\5\\7\\3\\4\\7\\2\\3\\3\\4\\1\\4\\1\\6\\6\\5\\
};
\addplot
table[row sep=\\,y index=0] {
data\\
5\\6\\5\\5\\7\\6\\7\\7\\5\\5\\7\\5\\6\\7\\6\\7\\7\\5\\5\\6\\
};

\addplot
table[row sep=\\,y index=0] {
data\\
5\\4\\6\\6\\3\\3\\7\\7\\6\\\\5\\4\\5\\6\\6\\6\\3\\6\\6\\6\\
};
\addplot
table[row sep=\\,y index=0] {
data\\
5\\4\\6\\5\\3\\3\\6\\5\\6\\2\\5\\4\\5\\5\\6\\5\\3\\3\\6\\7\\
};

\addplot
table[row sep=\\,y index=0] {
data\\
5\\4\\5\\3\\6\\5\\4\\6\\5\\5\\4\\4\\6\\6\\3\\5\\5\\6\\6\\6\\
};
\addplot
table[row sep=\\,y index=0] {
data\\
5\\5\\6\\6\\6\\5\\7\\5\\5\\4\\6\\5\\5\\4\\6\\6\\4\\4\\6\\6\\
};

\addplot
table[row sep=\\,y index=0] {
data\\
4\\3\\3\\3\\2\\2\\2\\6\\3\\5\\3\\3\\3\\5\\1\\3\\3\\5\\6\\4\\
};
\addplot
table[row sep=\\,y index=0] {
data\\
5\\5\\7\\7\\6\\5\\7\\5\\6\\6\\5\\5\\6\\3\\6\\7\\5\\5\\6\\5\\
};

\addplot
table[row sep=\\,y index=0] {
data\\
5\\4\\4\\4\\4\\3\\1\\3\\4\\6\\4\\3\\4\\3\\3\\3\\4\\4\\4\\4\\
};
\addplot
table[row sep=\\,y index=0] {
data\\
6\\6\\6\\7\\7\\4\\7\\7\\6\\3\\4\\5\\7\\5\\7\\6\\5\\4\\7\\5\\
};

\addplot
table[row sep=\\,y index=0] {
data\\
5\\3\\5\\6\\4\\5\\4\\6\\3\\7\\5\\5\\6\\4\\4\\6\\5\\5\\4\\5\\
};
\addplot
table[row sep=\\,y index=0] {
data\\
6\\6\\6\\6\\5\\3\\6\\7\\5\\4\\6\\5\\5\\3\\6\\6\\4\\5\\7\\5\\
};

\addplot
table[row sep=\\,y index=0] {
data\\
3\\3\\3\\3\\3\\3\\1\\6\\2\\2\\3\\4\\3\\3\\2\\3\\5\\5\\6\\3\\
};
\addplot
table[row sep=\\,y index=0] {
data\\
5\\5\\6\\7\\6\\2\\7\\5\\5\\5\\6\\5\\6\\3\\7\\7\\4\\5\\5\\4\\
};

\addplot
table[row sep=\\,y index=0] {
data\\
3\\5\\3\\4\\4\\5\\1\\7\\3\\4\\6\\3\\3\\2\\2\\5\\3\\5\\5\\4\\
};
\addplot
table[row sep=\\,y index=0] {
data\\
6\\6\\6\\7\\7\\6\\7\\7\\6\\7\\7\\6\\7\\7\\5\\7\\5\\5\\7\\6\\
};

\addplot
table[row sep=\\,y index=0] {
data\\
4\\3\\5\\4\\5\\\\5\\7\\4\\7\\5\\5\\6\\3\\4\\6\\6\\5\\6\\4\\
};
\addplot
table[row sep=\\,y index=0] {
data\\
4\\4\\5\\5\\4\\5\\6\\4\\5\\4\\4\\4\\4\\4\\4\\6\\3\\4\\4\\6\\
};

\legend{Video,VR}

\end{axis}

\draw[-latex] ($(plot.south west)+(0.3cm,0.2cm)$) -- ($(plot.south west)+(1.4cm,0.2cm)$)  node[midway,above,font=\scriptsize,inner sep=1pt] {better};

\end{tikzpicture}
}%
    \hfill
    \adjustbox{valign=T}{\includegraphics[height=5cm,clip,trim=100 20 140 0]{figures/mario_in_action.jpg}}
    \vspace{-.2cm}
    \caption{User study.
    Left: Statistical results, \ie median, lower and upper quartile (includes interquartile range), lower and upper fence, outliers (marked with $\bullet$) as well as the average value (marked with $\times$), for each aspect as recorded in our questionnaire. For most aspects, our VR-based system achieved higher ratings on the 7-point Likert scale than a video-based approach.
	Right: Our robot Mario in the most difficult part of the course.}
	\label{fig:study_questionnaire}
	\label{fig:mario_in_maze}
\end{figure*}

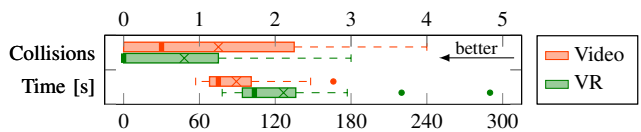
\begin{figure}
 \centering
﻿\begin{tikzpicture}[font = \footnotesize, every mark/.append style={mark size=1pt}]
\begin{axis}[
    name=plot,
    boxplot/draw direction=x,
    width=0.825\linewidth,
    height=2.5cm,
    boxplot={
        draw position={1.5 - 0.325/2 + 1.0*floor((\plotnumofactualtype + 0.001)/2) + 0.325*mod((\plotnumofactualtype + 0.001),2)},
        box extend=0.275,
        average=auto,
        every average/.style={/tikz/mark=x, mark size=2.5, mark options=black},
        every box/.style={draw, line width=0.5pt, fill=.!40!white},
        every median/.style={line width=2.0pt},
        every whisker/.style={dashed},
    },
    ymin=1,
    ymax=3,
    y dir=reverse,
    ytick={1,2,...,3},
    y tick label as interval,
    yticklabels={
        Collisions,
        Time [s]
    },
    y tick label style={
        align=center
    },
    xmin=-0.25,
    xmax=5.25,
    xtick={0, 1 ,..., 5},
    xticklabels = {0, 60, ..., 300},
    cycle list={{orange!50!red,green!50!black}},
    y dir=reverse,
    legend image code/.code={
        \draw [#1, fill=.!40!white] (0cm,-1.5pt) rectangle (0.3cm,1.5pt);
    },
    legend style={
        anchor=north west,
        at={($(1.0,1.0)+(0.2cm,-0.0cm)$)},
    },
    legend cell align={left},
    xticklabel pos = bottom,
]

\addplot
table[row sep=\\,y index=0] {
data\\
0\\4\\2\\3\\4\\0\\0\\1\\0\\0\\0\\3\\0\\0\\2\\1\\0\\3\\2\\0\\
};
\addplot
table[row sep=\\,y index=0] {
data\\
0\\2\\0\\0\\0\\0\\0\\1\\0\\0\\1\\1\\0\\2\\0\\0\\3\\2\\3\\1\\
};

\legend{Video,VR}

\end{axis}

\begin{axis}[
    name=plot,
    boxplot/draw direction=x,
    width=0.825\linewidth,
    height=2.5cm,
    boxplot={
        draw position={1.5 - 0.325/2 + 1.0*floor((\plotnumofactualtype + 2 + 0.001)/2) + 0.325*mod((\plotnumofactualtype + 2 + 0.001),2)},
        box extend=0.275,
        average=auto,
        every average/.style={/tikz/mark=x, mark size=2.5, mark options=black},
        every box/.style={draw, line width=0.5pt, fill=.!40!white},
        every median/.style={line width=2.0pt},
        every whisker/.style={dashed},
    },
    ymin=1,
    ymax=3,
    y dir=reverse,
    ytick={1,2,...,3},
    y tick label as interval,
    yticklabels={
        Collisions,
        Time [s]
    },
    y tick label style={
        align=center
    },
    xmin=-15,
    xmax=315,
    xtick={0, 60, ..., 300},
    xticklabels = {0, 1, ..., 5},
    cycle list={{orange!50!red,green!50!black}},
    y dir=reverse,
    legend image code/.code={
        \draw [#1, fill=.!40!white] (0cm,-1.5pt) rectangle (0.3cm,1.5pt);
    },
    legend style={
        anchor=north west,
        at={($(1.0,1.0)+(0.2cm,-0.0cm)$)},
    },
    legend cell align={left},
    xticklabel pos = bottom,
    axis x line* = top,
]

\addplot
table[row sep=\\,y index=0] {
data\\
68\\166\\113\\99\\68\\107\\140\\62\\73\\84\\69\\148\\65\\77\\96\\72\\59\\93\\57\\69\\
};
\addplot
table[row sep=\\,y index=0] {
data\\
100\\146\\220\\165\\133\\89\\290\\102\\93\\94\\120\\123\\78 \\105\\97\\94\\95\\177\\130\\80\\
};

\legend{Video,VR}

\end{axis}

\draw[latex-] ($(plot.north east)+(-1.1cm,-0.3cm)$) -- ($(plot.north east)+(-0.1cm,-0.3cm)$)
 node[midway,above,font=\scriptsize,inner sep=1pt] {better};

\end{tikzpicture}
  \vspace{-.2cm}
 \caption{User study: Statistical results for the number of collisions between robot and environment, and time needed for completing the course.}
 \label{fig:study_collision_time}
\end{figure}

After evaluating our VR-based teleoperation system in the scope of a user study, we provide a brief performance evaluation of the proposed approach as well as some proof-of-concept applications regarding how a remote user can interact with the scene.
A subset of this functionality is also demonstrated in the supplemental video.

\subsection{Implementation}

To implement the live teleoperation system, we use a laptop running the reconstruction client as well as the server component and a desktop computer that acts as the exploration client.
The laptop and the desktop computer have been equipped with an Intel Core i7-8700K CPU (laptop) and Intel Core i7-4930K CPU (desktop), 32\,GB RAM as well as a NVIDIA GTX 1080 GPU with 8\,GB VRAM.
Note that the system also allows additional exploration clients to be added if desired.
Additionally, the visualization of the data for the exploration client users is performed using an HTC Vive HMD device that has a native resolution of 1080$\times$1200 pixels per eye.
Due to the lens distortion applied by the HTC Vive system, the rendering resolution is 1512$\times$1680 pixels per eye as reported by the VR driver resulting in a total resolution of 3024$\times$1680 pixels.
Throughout all experiments, both computers were connected via WiFi.
Furthermore, we used a voxel resolution of 5\,mm and a truncation region of 60\,mm -- common choices for voxel-based 3D reconstruction.

\subsection{Evaluation of User Experience}

To assess the benefit of our immersive VR-based teleoperation system, we conducted a user study where we asked the participants to maneuver a robot through an elaborate course with challenges of different difficulties (see Fig.~\ref{fig:mario_in_maze}).
A reconstructed 3D model of the course is shown in Fig.~\ref{fig:teleop_reconstruction}.

\paragraph*{Participants}

In total, 20 participants voluntarily took part in the experiment (2 females and 18 males between 22 and 56 years, mean age 29.25 years).
All the participants were na\"ive to the goals of the experiment, provided informed consent, reported normal or corrected-to normal visual and hearing acuity.
Before conducting the experiments, the users got a brief training regarding the control instructions and a short practical training for all involved conditions.

\paragraph*{Stimuli}

Robot teleoperation was performed in two different modes (see Fig.~\ref{fig:teleop_experiment}).
In \emph{VR mode}, the robot navigation was performed based on immersing the user into the remote location of the robot via standard VR devices (in this case, the HTC Vive) and were able to follow the robot in terms of walking behind or, in case of larger distances, teleporting to the desired positions in the scene.
Here, the scene depicted in the HMD corresponds to the 3D model of the already reconstructed scene parts, which can be explored independently from the current view of the camera.
The rationale behind this experiment are the expected higher degrees of immersion and situation awareness as users get a better impression regarding distances in the scene as well as occurring obstacles.
Note that automatically following the robot instead is highly susceptible to motion sickness as it may not fit to the motion inherent to human behavior.
In \emph{video mode}, the users had to steer the robot through the same scenario purely based on video data depicting the current view of the camera on the robot arm.
Hereby, the flexibility of getting information outside the current camera view is lost.
As a consequence, we expect a lower situation awareness due to a more difficult perception of distances between objects in the scene as well as occurring obstacles.
Each participant performed the task once in VR and once in video mode.
We varied the order of these stimuli over the participants to avoid possibly occurring systematic bias due to training effects.
Since further multi-modal feedback is rather suited for attention purposes and less for accuracy of control, we left the integration and analysis of this aspect for future work.

\paragraph*{Performance measures}

In addition to gathering individual ratings for certain properties on a 7-point Likert scale, we also analyze the number of errors (collisions with the environment) made in the different modes and the total execution time required to navigate from the starting point to the target location.

\paragraph*{Discussion}

In Fig.~\ref{fig:study_questionnaire}, we show the statistical results obtained from the ratings provided by the participants for both VR-based and video-based robot teleoperation.
The main benefits of our VR system can be seen in the ratings regarding self-localization in the scene, maneuvering around narrow corners, avoiding obstacles, the assessment of the terrain for navigability as well as the ease of controlling the view.
For these aspects, the boxes defined by the medians and interquartile ranges do not overlap indicating a significant difference in favor of the VR-based teleoperation.
Furthermore, there is evidence that the VR mode is rated to be well-suited for teleoperation and that the robot can be easier moved to target positions.
These facts also support the general impression of the participants regarding a higher degree of situation awareness with the VR teleoperation, thereby following our expectations stated above.

On the other hand, it is likely that the higher degree of immersion also leads to closer, more-time consuming inspection, thus, limiting the speed of robot motion.
Furthermore, the perceived latency was rated slightly better for the video-based mode. The time until the scene data are streamed from the reconstruction client to the server, \ie the time until it is fully reconstructed or prefetched, depends on the camera movement and is within a few seconds.
A further slight deviation of the ratings in favor of the video-based mode can be seen regarding the resolution -- which is, in the case of the VR-based system, limited to the voxel resolution.
While the SLAMCast system supports on-demand local texture mapping of the current camera image onto the reconstructed 3D model, further advances towards the enhancement of texture resolution could help to bridge this last gap.

\begin{figure*}[t]
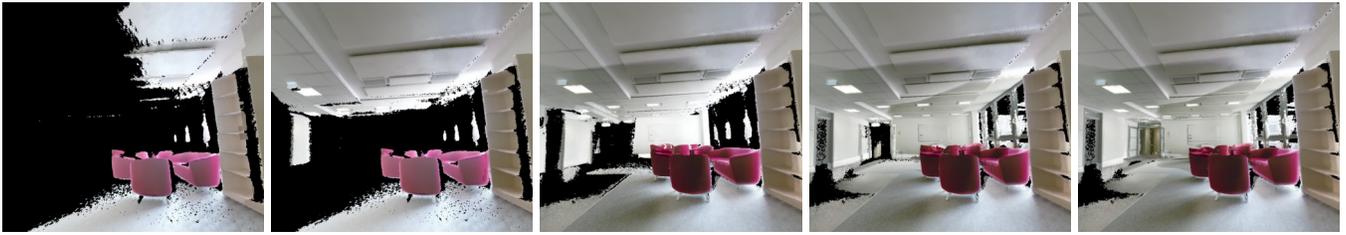

    \centering
    \includegraphics[width = 0.195\linewidth]{figures/1.png}
    \hfill
    \includegraphics[width = 0.195\linewidth]{figures/2.png}
    \hfill
    \includegraphics[width = 0.195\linewidth]{figures/4.png}
    \hfill
    \includegraphics[width = 0.195\linewidth]{figures/5.png}
    \hfill
    \includegraphics[width = 0.195\linewidth]{figures/6.png}
    \caption{Completion of scene model during capturing process: The images depict the scene model at different time steps. Depending on the regions that have been captured by the robot while moving through the scene, the captured 3D model of the environment gets more complete.}
    \label{fig:completeness_over_time}
\end{figure*}

Fig.~\ref{fig:study_collision_time} shows the statistical results for the number of collisions and time needed to complete the course with both modes.
The participants completed the course faster using video mode since more time was used in VR mode for inspecting the situation (\eg by walking around the robot in VR).
Teleportation inside the VR environment generally took some time, especially for participants without VR experience.
This could be improved by creating even more intuitive user interfaces for movement in VR and issuing navigation goals.
However, due to the improved situation awareness, more collisions could be avoided in VR mode.

\subsection{Performance Evaluation}

For performance evaluation, we first provide an overview of the bandwidth requirements as well as a visual validation of the completeness of the virtual 3D model generated over time of the proposed system.
For this purpose, we acquired two datasets based on the robotic platform and performed the reconstruction of the 3D models on the reconstruction client which are streamed to the server (first computer).
A benchmark client (second computer) requests voxel block data with a package size of 512 blocks at a fixed frame rate of 100\,Hz.
To avoid overheads that may bias the benchmark, we directly discard the received data.

We observed a mean bandwidth required for streaming the data from the server to the benchmark client of 14\,MBit/s and a maximum bandwidth of 25\,MBit/s, which is well within the typical limits of a standard Internet connection.
In Fig.~\ref{fig:completeness_over_time}, we demonstrate the completeness of the generated 3D model over time.
While at the beginning only a small area of the scene is visible to the exploration client, the remaining missing parts of the scene are progressively scanned by the robot, transmitted, and integrated in the client's local model.
In contrast to point cloud based techniques~\cite{Bruder:2014}, a closed-surface representation preserves the impression that objects or walls appear solid when viewed from varying distances.

\subsection{Interaction of Remote Users with the Scene}

\begin{figure}[t]
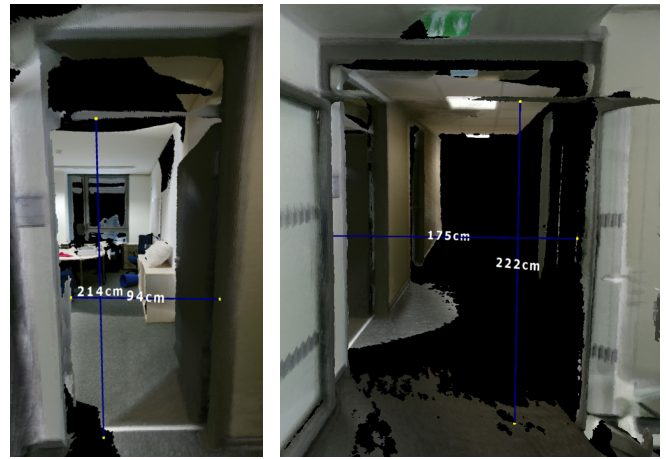

    \centering
    \includegraphics[width = .4\linewidth, height = .25\textheight, keepaspectratio]{figures/door_small.png}
    \hfill
    \includegraphics[width = .6\linewidth, height = .25\textheight, keepaspectratio]{figures/door_large.png}
    \caption{Examples of interactively taken measurements of heights and widths of a corridor as well as door widths taken to guide the further management process. The real sizes of the doors (\ie the ground truth values) are 95\,cm$\times$215\,cm (left) and 174\,cm$\times$222\,cm (right).}
    \label{fig:measurements}
\end{figure}

Managing contaminated site exploration or evacuation scenarios often involves the measurement of distances such as door widths in order to select and guide required equipment to the respective location.
For this purpose, we implemented operations for measuring 3D distances based on the controllers of the HMD device to allow user-scene interaction.
This can be useful in order to determine whether a different robot or the required equipment would fit through a narrow space, for example a door as shown in Fig.~\ref{fig:measurements}.
The measurement accuracy is determined by the voxel resolution, which is chosen according to the noise of the RGB-D camera as well as the tracking accuracy of the 3D reconstruction algorithm.
Considering the height and width of the doors measured in the corridor (see Fig.~\ref{fig:measurements}), we observed errors of up to 1\,cm which is sufficient for rescue management.

In addition, we also allow the remote user to label areas as interesting, suspicious or incomplete which is integrated into the overall map and the capturing robot may return to complete or refine the scan.
Since the SLAMCast system supports multi-client telepresence, a further remote user may perform this task while the other one is teleoperating the robot.
This enrichment of the captured 3D map with possibly annotated scene parts that have to be completed or refined can also directly be provided to further robots or the already used capturing robot.
Thereby the respective interactions of these robots with the scene can be guided (scan completion or refinement, transport of equipment).
So far, we did not include this functionality but leave it for future developments.

\section{Conclusion}

We presented a novel robot-based live immersive and teleoperation system for exploring contaminated places that are not accessible by humans.
For this purpose, we used a state-of-the-art robotic system which captures the environment with an RGB-D camera moved by its arm and transmits these data to a reconstruction and telepresence platform.
We demonstrated that our system allows interactive immersive scene exploration at acceptable bandwidth requirements as well as an immersive teleoperation experience.
Based on the implementation of several example operations, we also show the benefit of our proposed setup regarding the improvement of the degree of immersion and situation awareness for the precise navigation of the robot as well as the interactive measurement of objects within the scene.
In contrast, this level of immersion and interaction cannot be reached with video-only systems.

\addtolength{\textheight}{-0cm}   %

\bibliographystyle{IEEEtran}
\bibliography{literature}

\end{document}